\documentclass [twocolumn,epsfig,prl,10pt,floatfix,amsmath,amssym]{revtex4}
\usepackage[dvips]{epsfig}

\usepackage{float}

\begin{document}

\title{Continuous Crystallization in Hexagonally-Ordered Materials}
\author{Gregory M. Grason}
\affiliation{Department of Polymer Science and Engineering, University of Massachusetts, Amherst, MA 01003, USA}

\begin{abstract}
We demonstrate that the phase transition from columnar-hexagonal liquid crystal to hexagonal-crystalline solid falls into an unusual universality class, which in three dimensions allows for both discontinuous transitions as well as continuous transitions, characterized by a single set of exponents.  We show by a renormalization group calculation (to first order in $\epsilon = 4-d$) that the critical exponents of the continuous transition are precisely those of the XY model, giving rise to a continuous evolution of elastic moduli.  Although the fixed points of the present model are found to be identical to the XY model, the elastic compliance to deformations in the plane of hexagonal order, $\mu$, is nonetheless shown to critically influence the crystallization transition, with the continuous transition being driven to first order by fluctuations as the in-plane response grows weaker, $\mu \rightarrow 0$.    
  \end{abstract}
\date{\today}

\maketitle

On its way from an isotropic fluid to a three-dimensional (3D) crystal, a molecular system may pass through a nearly limitless sequence of mesophases of intermediate symmetry and bulk properties~\cite{pershan}.  For many systems close to crystallization, a liquid crystalline state with only a single continuous symmetry remaining appears, the columnar phase, in which molecular constituents adopt truly long-range two-dimensional (2D) order in the plane but remain fluid in the out-of-plane direction.  This ``minimally fluid" state is marginal to a full 3D solid in the sense that it is missing only a single elastic constant associated with shear along the columnar axis.  Historically, the columnar-hexagonal phase has been studied in the context of discotic, or disk-like, molecules ~\cite{safinya_liang_varady_clark, bushby_lozman}.  More recently, an interest in the complex ordering of stiff biopolymers, such as DNA~\cite{livolant_1, livolant_2, strey_etal1, clark}, and filamentous protein aggregates and viruses~\cite{spider_silk, grelet}, has refocused attention onto the specialized properties of columnar ordering and its role in biological assemblies.  Indeed, recent studies predict a direct connection between the type and quality of hexagonal ordering and the structural~\cite{grason_bruinsma_07} and mechanical~\cite{frey} properties of bundle aggregates of bio-filaments, such as those that constitute the cytoskeletal networks of living cells.

In this Letter, we analyze the critical properties of a system on the verge of losing the last vestige of fluid symmetry.   We predict that the columnar-to-solid transition may take place either 1) as a second-order transition characterized by the critical exponents of the XY model or 2) as a fluctuation-driven first-order transition, quite analogous to the finite temperature normal-to-superconductor transition in metals.  The nature of the transition is controlled by in-plane elastic response of the system, $\mu$, although curiously the coupling to in-plane modes of the system is not relevant to linear order at the fixed points of the renormalization group flow.  Below we give arguments to suggest that continuous crystallization can be associated with filamentous structure of molecular constituents, while a discontinuous transition will be more likely observed for disk-like molecules.  Finally, we summarize the universal ``critical elasticity" that emerges as a consequence of a continuous crystallization transition.

Dual to the columnar phase is the smectic-$A$ phase of rod-like molecules, which has broken translational symmetry along one axis and remains fluid in the 2D plane perpendicular to that axis.  Owing to the critical importance of fluctuations of ``one-dimensional solids"~\cite{anharmonic}, the nematic-to-smectic-$A$ transition is notoriously complex and has received considerable theoretical attention~\cite{halperin_lubensky_ma, halperin_lubensky, lubensky_chen, nelson_toner}.  In comparison, little is known about the statistical mechanics columnar-to-solid transition despite the fact that--as we will demonstrate--the critical properties of the former system may be considered a particular case of the present, more general, model describing the onset of one-dimensional periodic order.  Indeed, the statistical mechanics of the columnar-to-solid transition are surprisingly complex, leading to critical behavior with universal and non-universal aspects. 

A columnar liquid crystal is characterized by its elastic response to mechanical deformations.  Because it is translationally invariant along the columnar axis, a columnar system can be parameterized entirely by a two-component displacement field, ${\bf u}({\bf x})$, which describes the in-plane deviation of the column or filament backbones from their local equilibrium.  Assuming that columns are oriented along the $\hat{z}$ axis, the unit tangent vector representing backbone orientation is simply, $\hat{{\bf t}} = \hat{z}+\partial_z {\bf u}$.  Owing to the 2D hexagonal order, columnar systems resist in-plane mechanical strains like any isotropic 2D solid.   Hence, the Hamiltonian describing the elastic response of a columnar system has the following form \cite{degennes_prost},
\begin{multline}
\label{eq: Hperp}
{\cal H}_\perp = \frac{1}{2} \int d^3 x \big[ \lambda ~ u^2_{kk} + 2 \mu ~ u_{ij}u_{ij}  + K_1 ({\bf \nabla}_\perp \cdot \hat{{\bf t}})^2 \\ +K_2 ({\bf \nabla}_\perp \times \hat{{\bf t}})^2+ K_3 (\partial_z \hat{{\bf t}})^2 \big] .
\end{multline}
Here, $u_{ij} = (\partial_i u_j+\partial_j u_i)/2$ is the linearized 2D strain tensor, $\lambda$ and $\mu$ are the Lam\'e elastic parameters penalizing in-plane compression and shear, and $K_1$, $K_2$ and $K_3$ are the Frank elastic parameters for splay, twist and bend deformations, respectively.  While in a three-dimensional solid one may ignore the higher-order derivative contributions represented by the Frank elastic terms, in the columnar phase it is necessary to maintain at least $K_3\neq 0$ as the system is especially soft to bending deformations (i.e. $\partial_\perp {\bf u} =0$ but $\partial_z{\bf u}\neq 0$).  Such a system is stable to long-wavelength thermal fluctuations, and both orientational order and positional in-plane order are truly long-ranged in the thermodynamic limit.

The crucial distinction between 2D columnar vs. 3D crystalline order is the broken translational symmetry along the column axis of the latter.  We therefore construct a Landau-Ginzburg free energy in terms of a complex order parameter, $\psi({\bf x})$, representing broken translational order along the columnar direction.  Specifically, we may define it in terms of the density variation of the system near crystallization, $\delta \rho({\bf x}) \simeq \delta \rho_{{\rm Hex}}({\bf x})\big(1 + \psi({\bf x}) e^{ig z} + {\rm cc.}\big)$, where $2 \pi/g$ is the wavelength of the density modulation along the backbone direction.  Here, we focus on the simplest case, namely the transition from 2D hexagonal-columnar order to 3D hexagonal-solid order.  Clearly, $\psi({\bf x})$ represents a smectic-like ordering, with respective in-plane and out-of-plane gradients of the {\it phase} of $\psi$ corresponding to bending and compressive phonons of the nascent ordering.  As in smectics, rotational invariance of the system requires a specialized coupling of in-plane derivatives of $\psi$ to the displacements of column orientation.  With this in mind, the most general model one can write down (to second order in derivatives) to describe this transition has the following form,
\begin{multline}
\label{eq: H}
{\cal H} = \frac{1}{2} \int d^3 x \Big[ r |\psi|^2+\frac{u}{2} |\psi|^4 \\ + C_\perp \big| ({\bf \nabla}_\perp + i g  \hat{{\bf t}}_\perp) \psi \big|^2 + C_\parallel \big| (\partial_z + i g' u_{kk})\psi  \big|^2 \\ + \big( \delta \lambda ~  u^2_{kk}  +  2 \delta \mu ~ u_{ij}u_{ij} \big) |\psi|^2 \Big] .
\end{multline}
With the exception of the terms coupling to the in-plane strain tensor, this Hamiltonian has precisely the form required to describe the nematic-to-smectic $A$ transition~\cite{degennes_smectic}.  The critical properties of that model have been studied extensively, both in the context of liquid crystals~\cite{halperin_lubensky, lubensky_chen} as well as the normal-to-superconducting transition in metals~\cite{halperin_lubensky_ma}, and it was shown that as the critical temperature is approached ($r \rightarrow 0$) coupling of $\psi$ fluctuations to fluctuations of $\hat{{\bf t}}$ drive the transition weakly first order.  The origin of this effect can be traced to the fact for the nematic-to-smectic transitions, fluctuations of the director field are {\it massless}, especially soft, as their energy cost goes to zero roughly as $k^2| \hat{{\bf t}}_\perp({\bf k})|^2$.  In the present case, only bending modes are massless in this sense, as the in-plane ordering has quenched most of the long-wavelength fluctuations of the director field.  Hence, taking $\mu \rightarrow 0$ reduces the columnar phase to nematic so that this limit recovers the first-order nematic-to-smectic $A$ transition.  On the other hand, in the absence of out-of-plane positional order, orientational fluctuations coupling to order parameter fluctuations go to zero as $\langle |\hat{{\bf t}}_\perp|^2 \rangle \sim k_B T \mu^{-3/4} K_3^{-1/4}$ for large $\mu$~\cite{bruinsma_selinger}.   Intuitively, one would expect that in the limit that in-plane fluctuations are ``frozen out" by the stiff elastic response of the system, the phase transition is restored to second-order, as mean-field theory predicts.

To analyze the critical properties of this system more rigorously, we performed the renormalization group (RG) analysis of eqs. (\ref{eq: Hperp}) and (\ref{eq: H}) in $d=4-\epsilon$ dimensions.  For these purposes it is useful to rescale fields and dimensions to obtain a more isotropic gauge interaction with $C_\parallel = C_\perp = C$ and $g'= -g$.  In this case, fluctuations of $\psi$ couple to a transverse vector field, ${\bf n}$, with ${\bf n}_\perp = \partial_z {\bf u}$ and $n_z = -{\bf \nabla}_\perp \cdot {\bf u}$.  We choose to rescale lengths under RG transformation by $z'= e^{\delta \ell }z $ and ${\bf x}_\perp' = e^{\ell} {\bf x}_\perp $, where $\delta$ will be chosen to maintain the isotropic form of the gauge coupling.   At each RG step, we use standard diagrammatic methods, to calculate the perturbative corrections to the renormalized model from field fluctuations for large wavevectors $\Lambda e^{-\ell} < k < \Lambda$.  The full details of this analysis will be presented elsewhere.  It is nevertheless important to note here that we generically focus our analysis on the case when $K_i \Lambda^2 /\mu \ll 1 $ (i.e. small penetration depth).  It is straightforward to show from the rescaling procedure above that $\partial_\ell \ln [K_i\Lambda^2/\mu] = -2+4(1-\delta_\ell)+O(\epsilon)$.  A careful analysis shows that $1-\delta_\ell<0$ and generically flows to zero, hence all flows tend to the $K_i\Lambda^2/\mu \rightarrow 0$ limit.  

Although the analysis of this model involves many parameters, only a few parameters contribute to the fixed-point behavior.  Both $\delta \mu$ and $\delta \lambda$ constitute irrelevant parameters under RG flow, and we drop them from the subsequent analysis.  The most relevant couplings to the in-plane elastic response are captured by the parameters,
\begin{equation}
\label{eq: Q}
Q_\ell \equiv \frac{g_\ell^2}{12 \pi^2 \mu_\ell} \frac{1+\alpha_\ell}{2+\alpha_\ell} \Lambda^2 = -(1-\delta_\ell) ,
\end{equation}
and $4 \pi^2 R_\ell \equiv g_\ell^2/K_3$.  Here, $\alpha_\ell = \lambda_\ell/\mu_\ell$ and we have set $C_\ell =1$.  Given these definitions the RG flow is governed by the following recursion relations,
\begin{eqnarray}
\label{eq: rec}
\nonumber
\frac{d r_\ell}{d \ell} &\!\!=\!\!& \Big(2+\frac{2 \alpha_\ell+5}{\alpha_\ell+1} Q_\ell\Big) r_\ell+\frac{u_\ell \Lambda^2}{\pi^2}+3 Q_\ell  \frac{\alpha_\ell+3}{\alpha_\ell+1}\Lambda^2 \\
\nonumber \frac{d u
_\ell}{d \ell} &\!\!=\!\!&  \Big(\epsilon+\frac{5 \alpha_\ell+11}{\alpha_\ell+1} Q_\ell\Big) u_\ell-\frac{5 u_\ell^2}{4 \pi^2} \\ \nonumber && \ \ \ \ -2 \pi^2 R^{1/2}_\ell  (3 Q_\ell)^{3/2}\frac{1+(\alpha_\ell+2)^{3/2}}{(\alpha_\ell+1)^{3/2} } \\ \nonumber 
 \frac{d Q
_\ell}{d \ell} &\!\!=\!\!&-\big(2 - \epsilon+3 Q_\ell\big) Q_\ell \\
 \frac{d R
_\ell}{d \ell} &\!\!=\!\!& - \Big( - \epsilon - Q_\ell+ \frac{R_\ell}{6} \Big)R_\ell ,
\end{eqnarray}
with $d \alpha_\ell/d \ell = O(\epsilon^2)$.  Clearly, $Q_\ell \rightarrow 0$ for all flows near 4 dimensions, and therefore, $R_\ell$ flows to a stable fixed point, $R^*= 6 \epsilon$.  Therefore, the fixed points of eq. (\ref{eq: rec}) are identical to the isotropic XY model with a Gaussian fixed point, $r^*=u^*=0$, and a stable Wilson-Fisher fixed point, $r^*=-\epsilon \Lambda^2/5$ and $u^*= 4 \pi^2 \epsilon/5$. Indeed, to linear order, the couplings to in-plane elasticity are irrelevant at the fixed points, leaving the critical exponents of those points unchanged.  

\begin{figure}
\center \epsfig{file=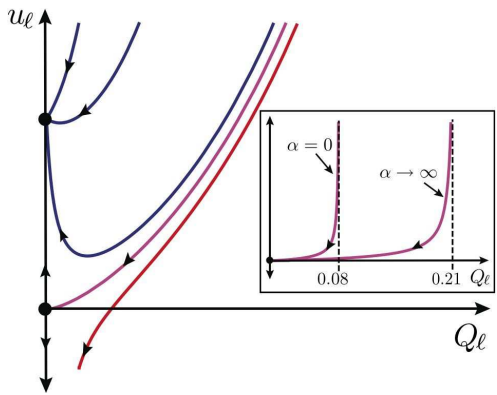, width=2.55in}\caption{The RG flows predicted by eq. \ref{eq: rec} for $\epsilon=1$, $\lambda=\alpha=0$ and $R_\ell$ at its stable fixed point.  The blue curves show parameter flow into the stable, Wilson-Fisher fixed point, while the red curve shows runaway flow.  Dividing stable and unstable flow regions is a separatrix (shown in purple) flowing into the Gaussian fixed point.  The inset show the seperatrix flow for both $\alpha=0$ and $\alpha \rightarrow \infty$ far from the fixed point.  According to these predictions, continuous crystallization is only possible for sufficiently small (large) values of $Q_0$ ($\mu_0$). } 
\label{fig: flows}
\end{figure}

In spite of the irrelevance of the gauge couplings at the fixed points, a very small non-linear perturbation, $O(\epsilon)$, around the Gaussian fixed point strongly alters the RG flow.  Focusing on the evolution of the quartic $\psi$ coupling in the vicinity of $R=R^*$, the coefficient of the $Q^{3/2}_\ell$ term is of order $\epsilon^{1/2}$, so that it generically dominates the flow in a neighborhood of order $\epsilon$ of that point.  In the limit of small ($\ll \epsilon$) $u_\ell$ and $Q_\ell$, the solution to the recursion relation for $u_\ell$ shows that RG flows are only stable provided that $u_0 >\epsilon C Q_0^{3/2}$, where $u_0$ and $Q_0$ are bare parameters and $ C $ is a non-universal constant, dependent on $\alpha$.  Outside of this region, flows runaway to $u_\ell \rightarrow - \infty$, which we interpret as a signature of a first-order transition.  Hence, this analysis predicts regions of stable and unstable RG flow for $d<4$.  In particular, we show numerical solutions to the flow equations $\epsilon =1$ in the $Q_\ell-u_\ell$ plane in Figure {\ref{fig: flows}} (evaluated for $R_\ell = R^*$).   A separatrix flowing into the Gaussian fixed point divides stable and runaway flows, indicating that the nature of transition, first or second order, is governed by the intial values $u_0$ and $Q_0$.  Interestingly, a similar RG flow is found for zero temperature, low-dimensional  superconducting systems in the presence of long-range repulsive interactions~\cite{fisher_grinstein}.   Not unlike the effect of in-plane order in columnar systems, the effect of long-range interactions is to sufficiently suppress critical gauge fluctuations of the superconducting system, allowing a phase transition to proceed as second order.

Interestingly, we see from Fig. {\ref{fig: flows}} that according to this analysis in $d=3$ stable RG flow is only possible below a maximum value of $Q$, which ranges from $0.08$ for $\alpha=0$ to $0.21$ for $\alpha \rightarrow \infty$.  Hence, this predicts a critical value of in-plane shear modulus, $\mu$, that separates first-order crystallization from second-order crystallization transitions.  Existing experimental studies of the columnar-to-solid transition confirm the existence of both continuous~\cite{livolant_3} and discontinuous~\cite{heiney_dejeu} transitions.  In order to understand what distinguishes the thermodynamics of these systems, we estimate the value of in-plane shear of a columnar system in terms of the microscopic in-plane, $a_\perp$, and out-of-plane, $a_\parallel$, length scales:  $\mu \sim a_\parallel^{-1} a_\perp^{-(d-1)}$ (in units of $k_B T$).  Here, $a_\parallel \simeq 2 \pi/g$ is the layering distance of smectic-like order along the column direction and $a_\perp$ corresponds to the in-plane separation of columns.  Using the eq. (\ref{eq: Q}) and the fact that our analysis is carried out near $d=4$, we estimate that the crystallization transition is first-order provided that $a_\parallel/a_\perp \lesssim (\Lambda a_\perp)^2$.  Based on this estimate, we expect that crystallization of high-aspect ratio, filament-like molecules will be continuous, while crystallization is discontinuous for low-aspect ratio, disk-like molecules.  Indeed, this picture is in perfect agreement with the respective results of DNA~\cite{livolant_3} and discotic molecule~\cite{heiney_dejeu} crystallization studies.  

These results establish that, to order $\epsilon$, the crystallization of columnar systems may occur either as a fluctuation-driven first-order transition or as a continuous, second-order transition characterized by XY exponents, i.e. $\xi \sim |t|^{-\nu}$ with $\nu \simeq 2/3$.  Here, $t$ is a generalized critical parameter, which may reflect either changes in temperature~\cite{heiney_dejeu} or concentration~\cite{grelet, livolant_3}.   We now briefly discuss the generic consequences of the continuous onset of broken translational symmetry in three dimensions, focusing on the response of the system to shear deformations that tend to slide the columns past one another.  This issue has been discussed previously in the context of a specific crystallization transition occurring in polyelectrolyte complexes, in which the elastic response to an external strain on the $d$-dimensional classical system was mapped onto the current response to an external vector potential, ${\bf A}_\perp$, of a zero-temperature superconducting system in $d-1$ dimensions ~\cite{grason_bruinsma}.

\begin{figure}
\center \epsfig{file=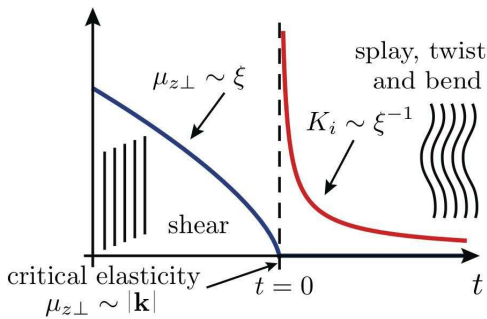, width=2.55in}\caption{A sketch of the singular elastic behavior of columnar systems undergoing continuous crystallization transitions.  The Frank elastic constants diverge at the critical point, while a response to uniform shear grows continuously with $t$.  At $t=0$ generic scaling arguments predict a non-analytic, critical elasticity.}
\label{fig: critical}
\end{figure}

In the most general case, the distinct properties of columnar and solid order are encoded in the elastic response $\mu_{z \perp} ({\bf k})$ to an inhomogeneous strain $u_{z \perp} ({\bf k})$.  Because a columnar system is translationally invariant, it has no response to uniform (${\bf k} =0$) strain.  In the solid phase, the long-wavelengh response to non-uniform displacements along the $\hat{z}$ direction has the form,
\begin{equation}
\delta {\cal H}_{\rm solid} = \frac{1}{2} \int \frac{ d^d k}{(2 \pi)^d} \mu_{z \perp} ({\bf k}) k_\perp^2 | \phi ({\bf k})|^2 ,
\end{equation}
where $\phi$ is the phase of the complex order parameter, $\psi$.  By dimensional analysis $\mu_{z \perp} ({\bf k} = 0)\sim \xi^{-(d-2)}$, since critical contributions to the free energy are governed entirely by the divergent correlation length.  Following a scaling argument of Fisher and coworkers developed for quantum phase transitions of two-dimensional superconductors~\cite{fisher_grinstein_girvin} we may deduce the wavelength dependence of $\mu_{z \perp} ({\bf k})$ near the critical point just below the crystallization transition.  Specifically, we must have $\mu_{z \perp} ({\bf k}) \sim \xi^{-(d-2)}f(k \xi)$, where $f(x)$ is a dimensionless function which goes to a constant for $x=0$ and must allow $\mu_{z \perp} ({\bf k})$ to remain finite as $x \rightarrow \infty$ since the response of the system to an inhomogenous ${\bf k} \neq 0$ shear will be finite at the critical point .  This requirement is satisfied by $f(x) = 1+ C_0 x^{(d-2)}$, which predicts an unusual {\it critical elastic response} to shear $\mu_{z \perp} ({\bf k})\sim |{\bf k}|$ in three dimensions when $\xi \rightarrow \infty$.   Approaching the critical point from the columnar phase where $\phi$ has no meaning, the singular contribution to the elastic energy of the system can be written as,
\begin{equation}
\delta {\cal H}_{\rm columnar}  = \frac{1}{2} \int \frac{ d^d k}{(2 \pi)^d} \mu_{z \perp} ({\bf k}) |\hat{{\bf t}} _\perp({\bf k})|^2 .
\end{equation}
Since the long-wavelength response of the columnar phase to orientational deformations goes as $k^2|\hat{{\bf t}} _\perp({\bf k})|^2$, we deduce that $f(x) \propto x^2$ above melting and $ \mu_{z \perp} ({\bf k}) \sim \xi^{-(d-4)} k^2$ (i.e. analogous to the {\it insulating} response of the quantum system).  This scaling predicts that in three dimensions renormalized Frank constants diverge as $\xi$ as the crystallization transition is approached from the columnar phase.  The singular dependence of elastic parameters for columnar systems undergoing a continuous crystallization transition is summarized in Figure \ref{fig: critical}.  

In summary, we carried out a first-order $\epsilon$ expansion below 4 dimensions to study the critical properties of the hexagonal-columnar liquid crystal to hexagonal crystal phase transition.  Depending on the in-plane elastic response of the system, this transition proceeds either as first order (driven by fluctuations) or second order described by the critical exponents of the XY model in the same dimension.  Finally, we argued that this rare example of a continuous crystallization transition, leads to a generic, singular evolution of bulk elastic properties.

\begin{acknowledgments}

I am indebted to R. Kamien and R. Bruinsma for many helpful discussions.  This work was supported by UMass, Amherst through a Healey Endowment Grant.  
\end{acknowledgments}

\end{document}